\documentclass[sigconf]{acmart}
\usepackage{algorithm}
\usepackage{algorithmic}
\usepackage{booktabs}
\usepackage{multirow}
\usepackage{amsmath}

\usepackage{amssymb}
\usepackage{enumitem}

\setcopyright{acmlicensed}
\copyrightyear{2018}
\acmYear{2018}
\acmDOI{XXXXXXX.XXXXXXX}
\acmConference[Conference acronym 'XX]{Make sure to enter the correct
  conference title from your rights confirmation email}{June 03--05,
  2018}{Woodstock, NY}
\acmISBN{978-1-4503-XXXX-X/2018/06}

\begin{document}

\title{Grevo: A Unified Generative Recommendation Framework with Evolutionary Item Indexing}

\author{Huanjie Wang}
\authornote{Both authors contributed equally to this research.}
\orcid{0009-0009-6046-2008}
\affiliation{%
    \institution{Beijing University of Posts and Telecommunications}
    \city{Beijing}
    \country{China}
}
\email{wanghuanjie@bupt.edu.cn}

\author{Liwei Guan}
\authornotemark[1]
\affiliation{%
    \institution{Beijing University of Posts and Telecommunications}
    \city{Beijing}
    \country{China}
}
\email{guanliwei@bupt.edu.cn}

\author{Zekai Sun}
\affiliation{%
    \institution{Beijing University of Posts and Telecommunications}
    \city{Beijing}
    \country{China}
}
\email{sunzekai@bupt.edu.cn}

\author{Hongwei Zhang}
\affiliation{%
    \institution{Beijing University of Posts and Telecommunications}
    \city{Beijing}
    \country{China}
}
\email{2232840184@bupt.edu.cn}

\author{Honghui Bao}
\authornote{Corresponding author.}
\affiliation{
  \institution{University of Illinois Chicago}
  \country{Chicago, USA}
}
\email{hbao@uic.edu}

\begin{abstract}
Generative recommendation has recently emerged as a promising paradigm that reformulates retrieval as autoregressive generation over semantic identifiers (SIDs), achieving strong performance and drawing increasing attention as an alternative to matching. Despite this progress, SIDs are typically frozen by a content-based tokenizer before the recommender is trained, leaving a persistent gap between what best reconstructs an item's content and what a recommender can predict from user behavior. Recent end-to-end methods close this gap by jointly training the tokenizer and the recommender, but coupling the two destabilizes the identifier space and requires a second learnable model, alignment losses, and usually a delicate alternating-optimization schedule. To address this issue, we propose \textbf{Grevo}, a unified \underline{G}enerative \underline{r}ecommendation framework with \underline{evo}lutionary item indexing, which treats the SID assignment itself as an evolvable discrete variable that adapts to behavioral feedback rather than as a tokenizer to be trained. Grevo builds on a single multitask recommender that unifies a behavioral SID generation task and a semantic SID grounding task, letting the recommender absorb the tokenizer's role. Through evolutionary item indexing, Grevo then uses the trained recommender itself as a posterior evaluator to reassign a budgeted set of high-risk identifiers under a fixed vocabulary and length. Together, these components turn index construction into a stable, feedback-driven search that requires no second learnable model, no alignment losses, and no alternating-optimization schedule. Extensive experiments on multiple real-world datasets demonstrate that Grevo consistently outperforms state-of-the-art generative recommendation methods.
\end{abstract}



\maketitle

\section{Introduction}
\label{sec:introduction}

\begin{figure*}[t]
\centering
\includegraphics[width=\textwidth]{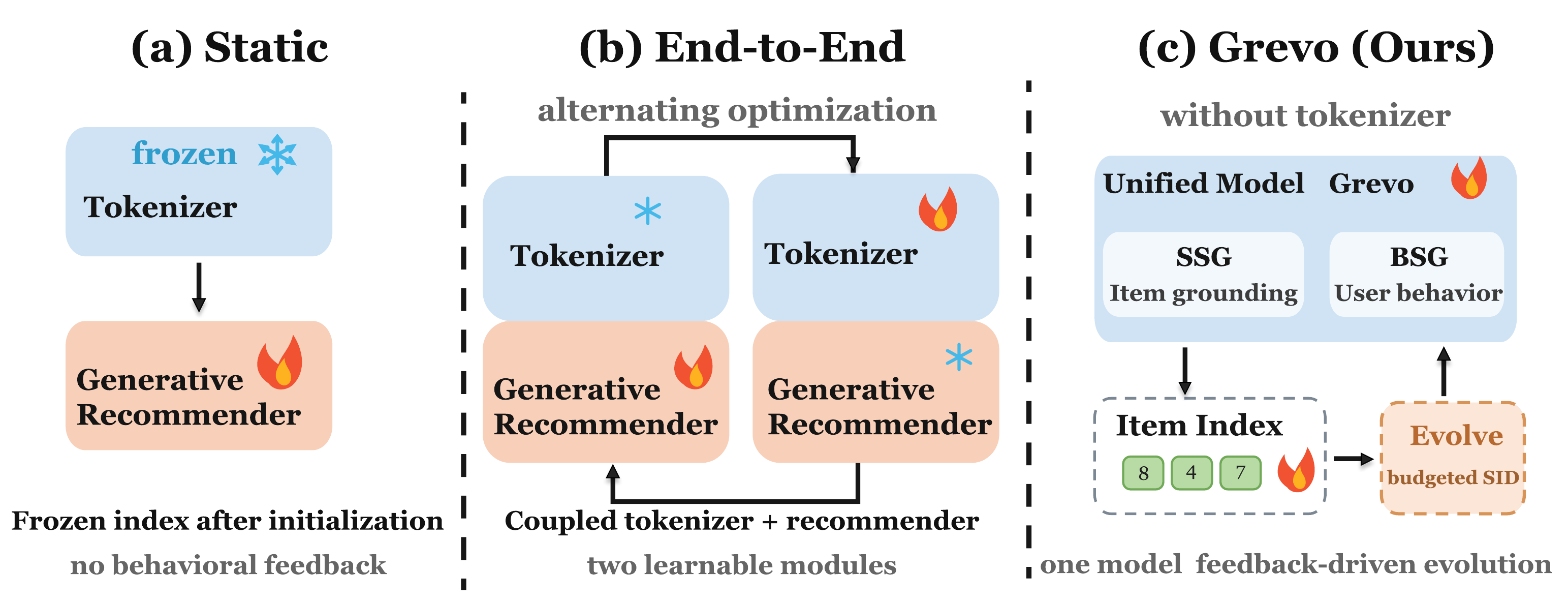}
\caption{Comparison of GR indexing paradigms. (a) Static methods freeze the tokenizer before recommender training, leaving no behavioral feedback path after initialization. (b) Coupled end-to-end methods jointly optimize a tokenizer and recommender through alternating updates and alignment losses. (c) Grevo keeps the initializer frozen, uses a unified BSG/SSG recommender as a posterior evaluator, and evolves a budgeted subset of SID assignments over rounds.}
\Description{Three side-by-side diagrams comparing GR indexing paradigms. Panel (a) shows a one-way arrow from a Tokenizer box to a Recommender box labeled Static, with no return path. Panel (b) shows bidirectional arrows between a Tokenizer and a Recommender connected by alignment losses and an alternating-optimization loop. Panel (c) shows Grevo: a frozen Initializer produces an initial SID index, which feeds a unified BSG/SSG Recommender; dashed arrows from the recommender indicate posterior evaluation that drives evolutionary SID reassignment back into the index across multiple rounds.}
\label{fig:overview}
\end{figure*}

Generative Recommendation (GR) has recently emerged as a promising paradigm, delivering strong empirical gains and establishing itself as a viable alternative to conventional matching-based retrieval~\cite{wang2024GenerativeRecommendationNextgeneration, li2024LargeLanguageModelsa, li2025SurveyGenerativeRecommendationa}. Rather than scoring a dense catalog, GR trains a single sequence-to-sequence~\cite{sutskever2014SequenceSequenceLearning} Transformer~\cite{vaswani2023AttentionAllYou} to emit the next item identifier token by token. This paradigm rests on two components: a \textit{tokenizer} that maps each item to a semantic identifier (SID) via codebook quantization~\cite{lee2022AutoregressiveImageGeneration} of its content embedding~\cite{rajput2023RecommenderSystemsGenerative, zhou2025OneRecTechnicalReport}, and a \textit{generative Transformer} trained to predict these identifiers from user behavior.

The design challenge centers on how these two components are coupled, and in particular on how SIDs are produced and updated. The dominant recipe is a decoupled two-stage pipeline~\cite{rajput2023RecommenderSystemsGenerative, zheng2024AdaptingLargeLanguage, wang2024EAGERTwoStreamGenerative, si2024GenerativeRetrievalSemantic} where a quantizer first maps each item to a static SID by minimizing content reconstruction error, and the recommender is then trained to predict these frozen identifiers. This separation induces a fundamental semantic--behavioral gap~\cite{wang2025LearnableItemTokenization, wang2024EAGERTwoStreamGenerative} since an SID optimized for content reconstruction is not necessarily aligned with the behavioral patterns that govern how users consume items, systematically capping downstream recommendation accuracy.

To bridge this gap, prior work falls into two broad categories. \textbf{Static methods} construct SIDs before recommender training and keep them frozen throughout, whether from content alone~\cite{rajput2023RecommenderSystemsGenerative, zheng2024AdaptingLargeLanguage} or with behavioral signals injected at construction time, as in LETTER~\cite{wang2025LearnableItemTokenization} and EAGER~\cite{wang2024EAGERTwoStreamGenerative}; once recommender training begins, the identifier space is fixed and no further correction is possible. \textbf{End-to-end coupled methods}~\cite{liu2025GenerativeRecommenderEndtoEnd, bai2025BiLevelOptimizationGenerative, wang2026pitdynamicpersonalizeditem} jointly optimize the tokenizer and recommender throughout training, which closes the feedback loop but requires a second learnable model, a staged alternating schedule, and careful stabilization to prevent the discrete targets from destabilizing recommender training. These methods collectively establish that behavioral signals should reshape the index, yet none enables post-hoc correction of identifiers without a coupled tokenizer or a second model.

Our approach starts from a simple observation that distinguishes GR from the language modeling it borrows its architecture from. In a language model the tokenizer is indispensable because raw strings must be segmented before the model can read them. In GR no such constraint exists because an item's SID is an index we are free to assign, so what GR requires is a well-organized index rather than a standing tokenizer that keeps producing one. This motivates treating the tokenizer as a disposable initializer that bootstraps the SID assignment once with semantic priors and is then discarded. To absorb the tokenizer's role, we fold both the recommendation view and the item grounding view into a single multitask recommender through two tasks distinguished only by an input prefix: \textit{Behavioral SID Generation} (BSG) and \textit{Semantic SID Grounding} (SSG). With the tokenizer's role subsumed, we then treat the SID assignment itself as an evolvable discrete variable, using the trained recommender's BSG/SSG posteriors as a surrogate signal to reassign a budgeted fraction of high-risk identifiers each round.

We instantiate this idea as \textbf{Grevo}, a unified generative recommendation framework built from two components: a \textit{unified multitask model} that jointly trains BSG and SSG, and \textit{evolutionary item indexing} that iteratively refines the SID assignment under a fixed vocabulary and length. Together they form a simple per-round lifecycle of pre-train, collect, enumerate, evolve and retrain, with no tokenizer involved beyond the initial bootstrap and no second model or alternating schedule required. Experiments on three Amazon benchmarks confirm that Grevo consistently outperforms strong generative baselines.

In summary, our main contributions are as follows:
\begin{itemize}[nosep,leftmargin=*]
\item We identify that GR requires a well-organized index rather than a standing tokenizer, and reframe behavior-aware identifier construction as evolutionary item indexing, absorbing the item$\rightarrow$SID role into the recommender itself via the SSG task.
\item We propose Grevo, combining a unified multitask model (BSG + SSG) with a budgeted evolutionary item indexing mechanism, requiring no second learnable model and no coupled optimization loop.
\item We conduct extensive experiments on three Amazon benchmarks with in-depth investigation, validating that Grevo consistently outperforms state-of-the-art generative recommendation methods.
\end{itemize}

\section{Related Work}
\label{sec:related work}

\noindent \textbf{Generative Recommendation.}
Generative Recommendation reformulates retrieval from vector similarity search~\cite{johnson2017BillionscaleSimilaritySearch, malkov2018EfficientRobustApproximate} into an autoregressive sequence-to-sequence task~\cite{sutskever2014SequenceSequenceLearning, wang2024GenerativeRecommendationNextgeneration, li2024LargeLanguageModelsa, li2025SurveyGenerativeRecommendationa}. This departs from traditional sequential recommenders~\cite{tang2018PersonalizedTopN, hidasi2018RecurrentNeuralNetworks, kang2018SelfAttentiveSequentialRecommendation, sun2019BERT4Rec, ma2019HierarchicalGatingNetworks, tang2025ThinkBeforeRecommend} that score a dense catalog, spanning convolutional, recurrent, and self-attentive architectures. Early generative work such as P5~\cite{geng2023RecommendationLanguageProcessing} predicts the next item with text-based tokens, while TIGER~\cite{rajput2023RecommenderSystemsGenerative} introduces semantic SIDs that represent items through discrete tokens quantized from content embeddings, allowing the model to capture item-level semantics and share knowledge across related items. Industrial systems such as OneRec~\cite{zhou2025OneRecTechnicalReport,zhou2025OneRecV2TechnicalReport} demonstrate the scalability of GR by deploying unified SIDs on large commercial infrastructures. LC-Rec~\cite{zheng2024AdaptingLargeLanguage} and OneReason~\cite{onerecteam2026onereasontechnicalreport} integrate SIDs into a large language model as part of its native vocabulary. A persistent obstacle to such deployments is the volatility of collaborative signals under real-time interaction, which drives severe distribution shifts in production streams~\cite{yang2023GenericLearningFramework, zou2025SurveyRealWorld}. Across these methods, recommendation quality remains tightly bound to the quality of the SIDs used as generation targets, motivating a growing interest in how identifiers are constructed and, as we argue, continually evolved.

\vspace{0.5em}
\noindent \textbf{Item Identifiers Assignment.}
Strategies for constructing item identifiers can be broadly categorized into static and end-to-end coupled approaches~\cite{wang2024GenerativeRecommendationNextgeneration, li2024LargeLanguageModelsa, li2025SurveyGenerativeRecommendationa}.
\textbf{Static methods} construct SIDs before recommender training and keep them frozen throughout. Content-only variants such as TIGER~\cite{rajput2023RecommenderSystemsGenerative} and LC-Rec~\cite{zheng2024AdaptingLargeLanguage} tokenize items offline via RQ-VAE~\cite{lee2022AutoregressiveImageGeneration}, optimizing reconstruction rather than the recommendation objective; behaviorally distinct items can thereby receive similar identifiers---a \textit{semantic--behavioral gap} that limits downstream precision~\cite{wang2025LearnableItemTokenization}. Behavior-injected variants such as LETTER~\cite{wang2025LearnableItemTokenization} and EAGER~\cite{wang2024EAGERTwoStreamGenerative} incorporate collaborative signals during tokenizer construction, but identifiers remain frozen once recommender training begins, so no further correction is possible.
\textbf{End-to-end coupled methods} close the feedback loop entirely: ETEGRec~\cite{liu2025GenerativeRecommenderEndtoEnd} aligns the tokenizer and recommender via representation losses and alternating optimization, but requires a second learnable model and a carefully staged training loop. PIT~\cite{wang2026pitdynamicpersonalizeditem} co-evolves a tokenizer and recommender on streaming data, but incurs heavy pipeline overhead and suffers popularity bias under long-horizon streaming. BLOGER~\cite{bai2025BiLevelOptimizationGenerative} frames tokenization as bi-level optimization, but gradient conflicts between the two tasks require surgery to avoid collapse, and its meta-gradient is myopic by design.
These methods collectively establish that behavioral signals should reshape the index, yet none allows the index to be corrected after training without a coupled tokenizer or a second learnable model. 

Grevo addresses this gap with two components: a unified multitask model that jointly trains BSG and SSG, letting a single recommender absorb both the recommendation and item grounding roles, and evolutionary item indexing that refines the SID assignment post-hoc using posterior signals from that trained model, with no tokenizer needed in the loop, as illustrated in Figure~\ref{fig:overview}.

\section{Methodology}
\label{sec:methodology}

\begin{figure*}[t]
\centering
\includegraphics[width=1.0\textwidth]{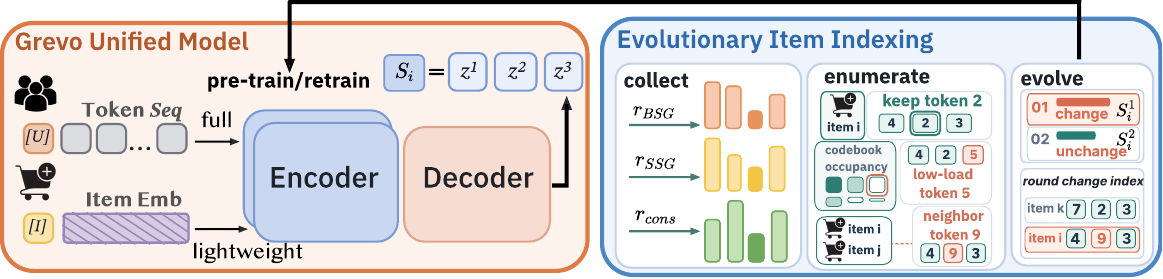}
\caption{The Grevo per-round lifecycle: \texttt{pre-train} $\rightarrow$ \texttt{collect} $\rightarrow$ \texttt{enumerate} $\rightarrow$ \texttt{evolve} $\rightarrow$ \texttt{retrain}. The trained BSG/SSG model acts as a posterior evaluator that scores candidate SID reassignments, and the evolved index feeds the next round. Only the SID assignment on mutable layers is updated.}
\Description{A circular diagram illustrating the five stages of the Grevo per-round lifecycle. Starting from pre-train, an arrow leads to collect, where the trained BSG/SSG model gathers posterior signals identifying high-risk items. A second arrow leads to enumerate, where constrained candidate SIDs are generated for each high-risk item. A third arrow leads to evolve, where candidates are scored by alignment gain and a budgeted subset of SID assignments is committed. A final arrow leads to retrain, which feeds the updated index back into pre-train for the next round. Mutable codebook layers are highlighted to indicate that only their token assignments change during evolution.}
\label{fig:lifecycle}
\end{figure*}

Grevo consists of two components. Section~\ref{sec:formulation} formulates the discrete SID evolution problem and introduces the \textit{unified multitask model} that lets a single recommender both recommend and ground items. The remainder of the section develops the second component, \textit{evolutionary item indexing}, which evolves the SID assignment under this model.

\subsection{Problem Formulation}
\label{sec:formulation}

Let $\mathcal{I}$ denote the item set. Each item $i \in \mathcal{I}$ is assigned a fixed-length SID of $K$ layers,
\begin{equation}
    s_i = (z_i^1, z_i^2, \ldots, z_i^K),
\end{equation}
where $z_i^k$ is the token emitted at codebook layer $k$, produced by an initial RQ-VAE~\cite{rajput2023RecommenderSystemsGenerative, lee2022AutoregressiveImageGeneration} or RQ-Kmeans~\cite{zhou2025OneRecTechnicalReport,zhou2025OneRecV2TechnicalReport} quantizer. Unlike end-to-end tokenizers, Grevo never opens the quantizer's parameters; it treats the assignment $\{s_i\}_{i\in\mathcal{I}}$ itself as an evolvable discrete variable and constrains its evolution to a restricted manifold. The key design principle is to preserve global index stability while allowing fine-grained local refinement. Because an evolved index respects the original length and vocabulary, it is a drop-in replacement for the initial index and requires no change to the model architecture.

\vspace{0.5em}
\noindent \textbf{Unified multitask model.}
The recommender $p_\theta$, instantiated as a sequence-to-sequence~\cite{sutskever2014SequenceSequenceLearning} Transformer~\cite{vaswani2023AttentionAllYou, raffel2020T5}, is pre-trained on two coupled tasks simultaneously over the current index. The two tasks share the same decoder but differ in their encoder input and BOS token:

\textbf{Behavioral SID Generation (BSG):} the generative recommendation task. The encoder receives a user's historical SID sequence $c$, prefixed with a \textit{user BOS token} $[\texttt{U}]$, and the decoder autoregressively generates the target item's SID $s$:
\begin{equation}
    p_\theta^{\mathrm{BSG}}(s \mid c) = \prod_{k=1}^{K} p_\theta\!\left(z^k \mid z^{<k},\, [\texttt{U}],\, c\right).
\end{equation}

\textbf{Semantic SID Grounding (SSG):} the item identification task. The encoder receives item $i$'s semantic representation $e_i$, prefixed with an \textit{item BOS token} $[\texttt{I}]$, and the decoder autoregressively generates the item's own SID $s_i$:
\begin{equation}
    p_\theta^{\mathrm{SSG}}(s \mid e_i) = \prod_{k=1}^{K} p_\theta\!\left(z^k \mid z^{<k},\, [\texttt{I}],\, e_i\right).
\end{equation}
The two tasks share the token embedding layer, and both are at heart the same problem---generating a codebook token path, so optimizing SSG directly improves the quality of the token representations the model generates and thereby lifts BSG recommendation accuracy, even when the index itself is held fixed. Crucially, SSG activates only the shallow layers of the model close to the token embeddings, whereas BSG engages the full model depth; this asymmetry keeps the additional cost of SSG negligible while still enriching the shared token space. The goal of Grevo is to evolve an SID assignment that is jointly favorable under both tasks: identifiers the model can generate from user behavior and recover from item semantics.

\vspace{0.5em}
\noindent \textbf{Evolutionary item indexing.}
The second component of Grevo pursues this goal by evolutionary item indexing: using the trained multitask model as a posterior evaluator, it repeatedly diagnoses high-risk identifiers and reassigns a budgeted fraction of them under a fixed vocabulary and fixed length, so that the index itself evolves toward the recommendation objective without ever reopening the tokenizer. The remainder of this section develops it as four interlocking parts: the round-based lifecycle that structures each evolution step (Section~\ref{sec:lifecycle}), the posterior signals collected from a trained recommender (Section~\ref{sec:signals}), the risk-based candidate generator (Section~\ref{sec:candidates}), and the alignment-gain scoring with budgeted commitment (Section~\ref{sec:scoring}).

\subsection{The Grevo Lifecycle}
\label{sec:lifecycle}

Grevo operates as a round-based loop shown in Figure~\ref{fig:lifecycle} where each round trains a recommender on the current index, evolves a small fraction of identifiers based on what the model has learned, and passes the improved index to the next round~\cite{bengio2009CurriculumLearning}. Each round begins by pre-training the unified BSG/SSG model on the current index, then collecting posterior signals to identify high-risk items. Candidate SIDs are enumerated and scored within the fixed-vocabulary, fixed-length manifold, and the highest-gain reassignments are committed under a budget to produce the next round's index. Conceptually, evolutionary item indexing is a training-after discrete refinement of the index: instead of back-propagating gradients into a tokenizer, it uses posterior diagnostics as a surrogate signal and takes a coordinate-descent-like step in the discrete SID space, evolving a small fraction of identifiers per round toward the recommendation objective. The evolve step of evolutionary item indexing consists of three stages, which collect items, enumerate them, and score and commit them, and these stages are executed between consecutive training rounds and are developed in Sections~\ref{sec:signals}--\ref{sec:scoring}.

\subsection{Collect: Posterior Signal Collection}
\label{sec:signals}

After each training round, Grevo reads two families of posterior signals from the model. Diagnostic signals identify which items the recommender struggles with most, measured by how far the model's predictions diverge from the target SIDs. Exact alignment signals then score each candidate SID $s$ for item $i$ by teacher-forcing it through both tasks: the BSG path score measures how easily the model generates the candidate from $C_i$, the set of user history sequences in which item $i$ appears as the target. Averaging over multiple contexts is inspired by ensemble learning~\cite{dietterich2000EnsembleMethodsMachine} and self-consistency~\cite{wang2023SelfConsistencyImprovesChain}: a single context may be noisy, but the mean log-probability over $C_i$ provides a stable estimate of the SID's behavioral quality,
\begin{equation}
    r_{\mathrm{BSG}}(i, s) = \frac{1}{|C_i|}\sum_{c \in C_i} \log p_\theta(s \mid [\texttt{U}], c),
\end{equation}
and the SSG path score measures how well the candidate can be recovered from the item's semantic representation $e_i$,
\begin{equation}
    r_{\mathrm{SSG}}(i, s) = \log p_\theta(s \mid [\texttt{I}], e_i).
\end{equation}
To jointly reflect both tasks, we further define a cross-task token-path consistency score that rewards candidates on which the two tasks agree at the token level. For each codebook layer $k \in \{1,\ldots,K\}$, let $u_k$ and $v_k$ denote the per-layer token log-probability of candidate $s$ under BSG and SSG respectively. The consistency score is
\begin{equation}
    r_{\mathrm{cons}}(i, s) = \frac{1}{K}\sum_{k=1}^{K} \frac{u_k + v_k - |u_k - v_k|}{2},
\end{equation}
which is high when both tasks favor the token path ($u_k$ and $v_k$ large) and the two tasks agree ($|u_k - v_k|$ small). A strong candidate is easy for BSG to generate, easy for SSG to identify, and consistent between the two tasks at every layer.

\subsection{Enumerate: Constrained Candidate Generation}
\label{sec:candidates}

For each high-risk item, Grevo enumerates a small set of candidate SIDs by replacing tokens only on the mutable layers. Let $\mathcal{M}\subseteq\{1,\ldots,K\}$ denote the mutable layers and let $\mathcal{R}(i)$ be the set of items that appear as posterior confusion neighbors of item $i$. For each mutable layer $k$, Grevo constructs a layer-wise token pool
\begin{equation}
    \mathcal{A}_i^k =
    \{z_i^k\}
    \cup
    \{z_j^k \mid j \in \mathcal{R}(i)\}
    \cup
    \operatorname{LowLoad}_k,
\end{equation}
where the three terms correspond respectively to the item's current token as a no-change anchor, tokens borrowed from behaviorally confused neighbors, and lightly loaded tokens from the existing layer-$k$ codebook. The candidate set is then
\begin{equation}
    \mathcal{S}_i =
    \{s' \mid s'^k \in \mathcal{A}_i^k \ \text{if}\ k\in\mathcal{M},\
    s'^k=z_i^k \ \text{otherwise}\},
\end{equation}
where $s'$ denotes a candidate SID and $s'^k$ its token at layer $k$. This construction keeps every proposal on the original fixed-vocabulary, fixed-length index manifold: Grevo never creates new codebook tokens, never changes immutable layers, and caps the number of enumerated candidates per item in practice.

\subsection{Evolve: Scoring and Budgeted Commitment}
\label{sec:scoring}

Each candidate is scored by the gain it achieves in the three alignment scores relative to the item's current SID,
\begin{equation}
    \Delta r_{\bullet}(i,s') = r_{\bullet}(i,s') - r_{\bullet}(i,s_i),
    \quad \bullet \in \{\mathrm{BSG}, \mathrm{SSG}, \mathrm{cons}\}.
\end{equation}
The total alignment gain is a weighted combination,
\begin{equation}
    \mathrm{score}(i,s') = \Delta r_{\mathrm{BSG}} + \lambda_1\,\Delta r_{\mathrm{SSG}} + \lambda_2\,\Delta r_{\mathrm{cons}},
\end{equation}
where the behavioral gain $\Delta r_{\mathrm{BSG}}$ is the primary objective and thus carries unit weight, $\Delta r_{\mathrm{SSG}}$ prevents an SID from drifting away from the item's semantics, and $\Delta r_{\mathrm{cons}}$ enforces per-layer consistency between the two tasks.

The solver then commits reassignments greedily by alignment gain, subject to a hard budget $\mathrm{Budget} = \lfloor \rho\,|\mathcal{I}| \rfloor$ controlled by budget ratio $\rho$, which caps the total number of evolved identifiers per round and allows at most one change per item. This gradual evolution is by design: end-to-end tokenizer training likewise results in only a fraction of SID assignments changing between consecutive checkpoints, rather than reorganizing the entire index at once. Grevo mirrors this principle explicitly --- keeping each round's changes small ensures the model can adapt to the updated index before the next evolution step, preventing the SID distribution from drifting too far in a single round and maintaining global index stability.

\section{Experiments}
\label{sec:experiments}

We conduct experiments to answer the following research questions:
\begin{itemize}[label=$\bullet$, leftmargin=*]
    \item \textbf{RQ1:} How does Grevo perform against representative traditional sequential and generative recommendation baselines?
    \item \textbf{RQ2:}  Do Grevo's design components each contribute to the gain, and do the evolved identifiers remain more decodable on held-out user contexts?
    \item \textbf{RQ3:} How does Grevo behave across mutable-layer configurations, evolution budgets, alignment weights, and evolution rounds?
\end{itemize}

\subsection{Experimental Setup}
\label{sec:setup}

\noindent \textbf{Datasets.} We evaluate on three widely used datasets from the Amazon Review 2014~\cite{he2016UpsDownsModeling} collection: ``Beauty'', ``Sports and Outdoors'', and ``Toys and Games''. Following established preprocessing protocols~\cite{kang2018SelfAttentiveSequentialRecommendation,wang2025LearnableItemTokenization}, we treat all ratings as positive implicit feedback and adopt the 5-core setting~\cite{kang2018SelfAttentiveSequentialRecommendation}, filtering out users and items with fewer than five interactions. User behavior sequences are constructed chronologically with a maximum historical sequence length of $50$. We employ the widely used leave-one-out strategy~\cite{kang2018SelfAttentiveSequentialRecommendation}, using the last interaction for testing, the second-to-last for validation, and the remainder for training. The statistics of the processed datasets are summarized in Table~\ref{tab:dataset}.

\begin{table}[t]
\centering
\caption{Statistics of the preprocessed datasets.}
\label{tab:dataset}
\resizebox{\columnwidth}{!}{%
\begin{tabular}{lccccc}
\toprule
Dataset & \#Users & \#Items & \#Interactions & Sparsity & Avg. Length \\
\midrule
\textbf{Beauty} & 22,363 & 12,101 & 198,502 & 99.93\% & 8.9 \\
\textbf{Sports} & 35,598 & 18,357 & 296,337 & 99.95\% & 8.3 \\
\textbf{Toys}   & 19,412 & 11,924 & 167,597 & 99.93\% & 8.6 \\
\bottomrule
\end{tabular}}
\end{table}

\noindent \textbf{Baselines.} We compare Grevo against a wide range of competitive baselines, grouped into traditional sequential recommenders and recent generative recommenders.

\begin{table*}[t]
\centering
\caption{Overall performance across three Amazon datasets. Grevo is instantiated on three generative backbones. For TIGER and LETTER we report the plain backbone, \textbf{+Grevo (no-evo)} (the unified multitask model with BSG+SSG on the initial index, no evolution), and \textbf{+Grevo} (with evolutionary item indexing added on top); for LC-Rec we report the plain backbone and \textbf{+Grevo}. Per backbone group, \textbf{bold} marks the best result, each compared only against the baselines above.}
\label{tab:baselines}
\setlength{\tabcolsep}{4pt}
\resizebox{\textwidth}{!}{%
\begin{tabular}{l|cccc|cccc|cccc}
\toprule
\multirow{2}{*}{\textbf{Model}} & \multicolumn{4}{c|}{\textbf{Beauty}} & \multicolumn{4}{c|}{\textbf{Sports}} & \multicolumn{4}{c}{\textbf{Toys}} \\
\cmidrule(lr){2-5} \cmidrule(lr){6-9} \cmidrule(lr){10-13}
 & R@5 & R@10 & N@5 & N@10 & R@5 & R@10 & N@5 & N@10 & R@5 & R@10 & N@5 & N@10 \\
\midrule
\textbf{HGN}      & 0.0337 & 0.0554 & 0.0207 & 0.0277 & 0.0192 & 0.0325 & 0.0117 & 0.0160 & 0.0356 & 0.0565 & 0.0216 & 0.0284 \\
\textbf{GRU4Rec}  & 0.0401 & 0.0614 & 0.0260 & 0.0329 & 0.0194 & 0.0320 & 0.0124 & 0.0164 & 0.0358 & 0.0571 & 0.0234 & 0.0302 \\
\textbf{BERT4Rec} & 0.0226 & 0.0385 & 0.0140 & 0.0191 & 0.0102 & 0.0168 & 0.0063 & 0.0084 & 0.0194 & 0.0302 & 0.0120 & 0.0155 \\
\textbf{SASRec}   & 0.0403 & 0.0592 & 0.0266 & 0.0327 & 0.0203 & 0.0320 & 0.0133 & 0.0171 & 0.0424 & 0.0609 & 0.0282 & 0.0342 \\
\textbf{HSTU}     & 0.0423 & 0.0623 & 0.0282 & 0.0347 & 0.0236 & 0.0348 & 0.0157 & 0.0193 & 0.0405 & 0.0582 & 0.0283 & 0.0339 \\
\textbf{OneRec-V2} & 0.0413 & 0.0628 & 0.0277 & 0.0346 & 0.0240 & 0.0402 & 0.0147 & 0.0199 & 0.0361 & 0.0581 & 0.0223 & 0.0294 \\
\textbf{ETEGRec}  & 0.0408 & 0.0657 & 0.0270 & 0.0344 & 0.0269 & 0.0435 & 0.0164 & 0.0221 & 0.0391 & 0.0602 & 0.0245 & 0.0321 \\
\textbf{BLOGER}   & 0.0344 & 0.0546 & 0.0220 & 0.0285 & 0.0259 & 0.0439 & 0.0165 & 0.0223 & 0.0348 & 0.0543  & 0.0221 & 0.0285 \\
\midrule
\textbf{LC-Rec}   & 0.0433 & 0.0653 & 0.0289 & 0.0360 & 0.0271 & 0.0428 & 0.0177 & 0.0228 & 0.0466 & 0.0687 & 0.0323 & 0.0394 \\
\textbf{\;\;+Grevo}   & \textbf{0.0523} & \textbf{0.0777} & \textbf{0.0335} & \textbf{0.0428} & \textbf{0.0278} & \textbf{0.0441} & \textbf{0.0180} & \textbf{0.0233} & \textbf{0.0489} & \textbf{0.0729} & \textbf{0.0333} & \textbf{0.0410} \\
\midrule
\textbf{TIGER}   & 0.0403 & 0.0648 & 0.0256 & 0.0336 & 0.0262 & 0.0418 & 0.0165 & 0.0215 & 0.0352 & 0.0562 &  0.0220 & 0.0287 \\
\textbf{\;\;+Grevo (no-evo)}   & 0.0428 & 0.0713 & 0.0265 & 0.0356 & 0.0283 & 0.0470 & 0.0179 & 0.0239 & 0.0404 & 0.0662 & 0.0254 & 0.0337 \\
\textbf{\;\;+Grevo}  & \textbf{0.0484} & \textbf{0.0778} & \textbf{0.0318} & \textbf{0.0412} & \textbf{0.0294} & \textbf{0.0489} & \textbf{0.0188} & \textbf{0.0250} & \textbf{0.0455} & \textbf{0.0704} & \textbf{0.0294} & \textbf{0.0375} \\
\midrule
\textbf{LETTER}  &  0.0411 & 0.0654 &  0.0263 & 0.0340 & 0.0266 & 0.0439 & 0.0167 & 0.0223 &  0.0383 & 0.0599 & 0.0239 &  0.0308 \\
\textbf{\;\;+Grevo (no-evo)}  & 0.0431 & 0.0707 & 0.0275 & 0.0364 & 0.0292 & 0.0479 & 0.0185 & 0.0245 & 0.0428 & 0.0692 & 0.0269 & 0.0353 \\
\textbf{\;\;+Grevo} & \textbf{0.0476} & \textbf{0.0751} & \textbf{0.0312} & \textbf{0.0401} & \textbf{0.0301} & \textbf{0.0483} & \textbf{0.0192} & \textbf{0.0251} & \textbf{0.0477} & \textbf{0.0720} & \textbf{0.0309} & \textbf{0.0388} \\
\bottomrule
\end{tabular}%
}
\end{table*}

\textit{Traditional sequential models:}
\textbf{HGN}~\cite{ma2019HierarchicalGatingNetworks} introduces hierarchical feature and instance gating to capture both short-term and long-term user interests;
\textbf{GRU4Rec}~\cite{hidasi2018RecurrentNeuralNetworks} is a pioneering session-based method that models user click behaviors with Gated Recurrent Units;
\textbf{BERT4Rec}~\cite{sun2019BERT4Rec} adapts a deep bidirectional Transformer with a Cloze objective to capture bidirectional context;
and \textbf{SASRec}~\cite{kang2018SelfAttentiveSequentialRecommendation} uses unidirectional self-attention to capture long-term sequential dependencies.

\textit{Generative recommendation models:}
\textbf{HSTU}~\cite{zhai2024ActionsSpeakLouder} is an optimized sequential architecture designed for large-scale industrial deployment;
\textbf{OneRec-V2}~\cite{zhou2025OneRecV2TechnicalReport} is a large-scale end-to-end generative recommender using RQ-Kmeans;
\textbf{TIGER}~\cite{rajput2023RecommenderSystemsGenerative} formulates recommendation as generative retrieval, using a hierarchical RQ-VAE to produce semantic identifiers;
\textbf{LETTER}~\cite{wang2025LearnableItemTokenization} optimizes item tokenization by injecting semantic, collaborative, and diversity regularization into a learnable framework;
\textbf{ETEGRec}~\cite{liu2025GenerativeRecommenderEndtoEnd} proposes an end-to-end training strategy with alternating optimization and auxiliary recommendation-oriented losses;
\textbf{LC-Rec}~\cite{zheng2024AdaptingLargeLanguage} adapts large language models for recommendation, integrating collaborative semantics into generation via vector quantization;
and \textbf{BLOGER}~\cite{bai2025BiLevelOptimizationGenerative} frames tokenization as a bi-level optimization, back-propagating the recommendation loss into the tokenizer with gradient surgery.

\noindent \textbf{Evaluation Metrics.} We adopt two standard metrics, \textbf{Recall@$K$} and \textbf{NDCG@$K$}, reporting results for $K \in \{5, 10\}$ following~\cite{rajput2023RecommenderSystemsGenerative} to analyze performance at different ranking positions.

\noindent \textbf{Implementation Details.} For a fair comparison, we uniformly employ a frozen LLaMA3-8B~\cite{grattafiori2024Llama3Herd} to extract item embeddings for codebook construction in TIGER~\cite{rajput2023RecommenderSystemsGenerative}, LETTER~\cite{wang2025LearnableItemTokenization}, LC-Rec~\cite{zheng2024AdaptingLargeLanguage}, and OneRec-V2~\cite{zhou2025OneRecV2TechnicalReport}. Specifically, for TIGER and LETTER, we adopt the open-source implementation provided by LETTER. For LC-Rec, while the item embeddings follow the standardized extraction process, the subsequent fine-tuning stage is conducted using LLaMA-7B~\cite{touvron2023LLaMAOpenEfficient}, consistent with its original architecture. TIGER, LETTER, and LC-Rec share an identical quantization configuration using RQ-VAE~\cite{lee2022AutoregressiveImageGeneration} with 4 codebooks, each containing 256 codes of dimension 32, whereas OneRec-V2 constructs semantic IDs using RQ-Kmeans~\cite{zhou2025OneRecTechnicalReport} based on the same extracted embeddings. For ETEGRec~\cite{liu2025GenerativeRecommenderEndtoEnd} and BLOGER~\cite{bai2025BiLevelOptimizationGenerative}, we utilize their official open-source implementations. Traditional recommendation baselines are implemented using the open-source recommendation framework RecBole~\cite{zhao2021RecBoleUnifiedComprehensive}. Grevo is implemented based on a T5-style~\cite{raffel2020T5} encoder-decoder architecture shared by BSG and SSG. By default, the model consists of 8 encoder and 8 decoder layers with a hidden size of 256, 4 attention heads, a 2048-dim feed-forward network with Swish~\cite{ramachandran2018searching} activation, and a dropout rate of 0.2. Each round is trained using seed 32 for up to 75 epochs with early stopping, SSG loss weight 0.07, and maximum sequence length of 50. For evolutionary indexing, the SID length and vocabulary size remain fixed where evolve round is 1. Posterior diagnostics employ constrained beam search~\cite{sutskever2014SequenceSequenceLearning} with a beam size of 50, retaining up to 32 user contexts per target item for exact alignment scoring. In the default evolutionary setup, Grevo edits the last two SID layers, enumerates up to 64 candidates per item from confusion-neighbor and low-load token sources, and sets $\rho=0.05$, $\lambda_1=0.25$, and $\lambda_2=0.15$. Ablation and parameter studies vary one factor at a time on Beauty with TIGER SID.

\subsection{Overall Performance (RQ1)}
\label{sec:main-results}

Table~\ref{tab:baselines} reports Recall (R) and NDCG (N) at $K\in\{5,10\}$. We evaluate Grevo natively on the TIGER and LETTER backbones: each group contains the plain backbone, +Grevo (no-evo), which trains the unified BSG+SSG model on the fixed initial index, and the full +Grevo model. We additionally conduct an index-transfer test with LC-Rec: its+Grevo row is trained directly with the evolved SID produced by TIGER+Grevo, without applying Grevo's BSG+SSG training or a tokenizer optimization to LC-Rec itself. Based on the results, we make the following observations:
\begin{itemize}[label=$\bullet$, leftmargin=*]
    \item Grevo consistently achieves the strongest overall performance across datasets and metrics. Its advantage comes from treating the item index as an adaptive part of generative recommendation rather than a fixed preprocessing artifact. The unified BSG+SSG model jointly aligns behavioral SID generation with semantic grounding through shared token representations, while the evolutionary step uses posterior feedback from both tasks to repair high-risk identifiers under a fixed vocabulary and length. This coupling yields identifiers that are simultaneously easier to generate from user behavior and better grounded in item semantics, providing a more suitable target space for generative recommendation.

    \item The gains from the unified model and evolutionary indexing are complementary. On both TIGER and LETTER, +Grevo (no-evo) already improves over the corresponding backbone, showing that jointly learning behavioral generation and semantic grounding benefits recommendation even when the initial SID is fixed. Adding evolutionary indexing then delivers a further improvement, confirming that the recommender's posterior feedback provides information that cannot be recovered from fixed-index multitask training alone. This conclusion is further supported by the LC-Rec transfer setting: LC-Rec uses only the evolved index obtained from TIGER+Grevo for training, without any LC-Rec-specific Grevo module. As the results show, its improvement isolates the effect of the index and shows that a well-evolved SID can be directly transferred to strengthen another generative recommender. In other words, generative recommendation crucially needs a good item index, rather than an additional backbone-specific learnable tokenizer.

    \item Traditional sequential recommenders, including HGN, GRU4Rec, BERT4Rec, and SASRec, generally underperform recent generative approaches. These methods predict over item-specific embeddings or categorical identifiers, which offer limited semantic structure and do not provide a hierarchical target space for generation. Recent generative recommenders alleviate this limitation by introducing semantic identifiers or collaborative semantic representations: TIGER and LETTER use residual-quantized identifiers, LC-Rec incorporates collaborative semantics into generation, and OneRec-V2 uses quantized semantic IDs. Their stronger results demonstrate the value of structured item representations. ETEGRec and BLOGER further advance this line of work by jointly learning the tokenizer and generative recommender, allowing identifier construction to become aware of the downstream recommendation task. Building on this insight, Grevo uses budgeted, posterior-guided discrete evolution to progressively refine the index with feedback from the trained recommender. As a result, It provides a direct and lightweight route to a better-aligned, recommendation-aware item index, without introducing a second learnable tokenizer. The results show that this design achieves stronger recommendation effectiveness without requiring a separately learned tokenizer.
\end{itemize}

\subsection{In-depth Analysis}
\label{sec:in-depth}

\subsubsection{Ablation Study (RQ2)}
\label{sec:ablations}

To investigate the individual contribution of Grevo's components, we conduct an ablation study on Amazon Beauty with TIGER SID. As summarized in Table~\ref{tab:ablation}, we compare the full model against variants that simplify the posterior scorer or remove one candidate-token source; all variants use the same mutable layers and evolution budget.
\begin{itemize}[label=$\bullet$, leftmargin=*]
    \item \textbf{$r_{\mathrm{SSG}}$ only}: Retains only the semantic SID-grounding score when selecting candidate identifiers, removing both the behavioral posterior and cross-task consistency. Its clear performance decline shows that semantic recoverability alone is insufficient for identifying the SIDs that best serve the recommendation task; the index must also reflect how users generate items from their behavioral histories.

    \item \textbf{$r_{\mathrm{BSG}}$ only}: Selects candidate SIDs solely according to the behavioral SID-generation posterior, without semantic grounding or consistency regularization. This variant performs substantially better than the SSG-only setting, validating that the behavioral posterior is the primary signal for repairing recommendation-critical identifiers. Its remaining gap to the full model further shows that behavior alone can favor identifiers that are insufficiently grounded or unstable across the two tasks.

    \item \textbf{$r_{\mathrm{BSG}}+r_{\mathrm{SSG}}$}: Combines behavioral and semantic posterior signals but removes the cross-task consistency term. The performance decrease relative to Grevo (full) demonstrates that encouraging both tasks to favor the same token path is beneficial: it filters candidates that look favorable to one task but are not jointly reliable as an item identifier.

    \item \textbf{w/o neighbor tokens}: Disables tokens borrowed from posterior confusion neighbors during constrained candidate generation, leaving only the current token and low-load alternatives. The pronounced degradation indicates that confused neighbors provide the most informative local repair directions, because they expose the behavioral ambiguities that the current index fails to distinguish.

    \item \textbf{w/o low-load tokens}: Removes lightly loaded codebook tokens from the candidate pool while retaining confusion-neighbor tokens. The resulting drop confirms that low-load tokens complement neighbor-derived proposals by supplying available codebook capacity for useful reassignment, rather than restricting evolution to the tokens already occupied by confusable items.
\end{itemize}

\begin{table}[t]
\centering
\caption{Ablation of Grevo's design choices on Amazon Beauty with TIGER SID. All variants use one evolution round with budget ratio $\rho=0.05$.}
\label{tab:ablation}

\setlength{\tabcolsep}{4pt}

\begin{tabular}{lcccc}
\toprule
\textbf{Variant} & R@5 & R@10 & N@5 & N@10 \\
\midrule
\textbf{Grevo (full)} 
& \textbf{0.0484} & \textbf{0.0778} & \textbf{0.0318} & \textbf{0.0412} \\
\midrule
\multicolumn{5}{l}{\textbf{Posterior Signal Collection}} \\
\; \textbf{$r_{\mathrm{SSG}}$ only} 
& 0.0453 & 0.0714 & 0.0291 & 0.0375 \\

\; \textbf{$r_{\mathrm{BSG}}$ only} 
& 0.0472 & 0.0760 & 0.0307 & 0.0401 \\

\; \textbf{$r_{\mathrm{BSG}}+r_{\mathrm{SSG}}$}
& 0.0477 & 0.0765 & 0.0313 & 0.0406 \\

\midrule
\multicolumn{5}{l}{\textbf{Candidate Token Sources}} \\

\; \textbf{w/o neighbor tokens}
& 0.0440 & 0.0706 & 0.0279 & 0.0364 \\

\; \textbf{w/o low-load tokens}
& 0.0479 & 0.0762 & 0.0308 & 0.0399 \\

\bottomrule
\end{tabular}

\end{table}

Overall, the results show that Grevo requires both a behavior-aware posterior scorer and a diverse but constrained candidate pool. The behavioral signal supplies the main recommendation objective, semantic grounding and consistency make the edits jointly reliable, and the two candidate sources provide complementary directions for repairing the index.

\begin{figure*}[t]
    \centering
    \includegraphics[width=1.0\textwidth]{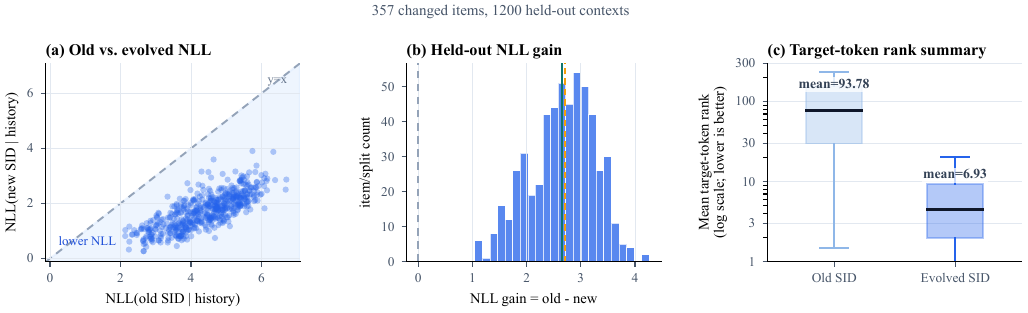}
    \caption{Held-out SID decodability on Beauty. (a) Paired NLLs of original and evolved SIDs; points below $y=x$ favor evolved SIDs. (b) NLL gain (old minus evolved, the green line shows the mean NLL gain, while the yellow line shows the median NLL gain.). (c) Correct-token rank distributions (lower is better), summarized by box plots. A logarithmic axis accommodates the two-order-of-magnitude rank range; ticks and means remain in original units.}
    \Description{Three panels compare evolved and original SIDs on held-out decodability: all per-edit NLL points lie below the equality line, NLL gains are positive, and log-scale box plots show substantially lower target-token ranks for evolved SIDs.}
    \label{fig:heldout-dec}
\end{figure*}

\subsubsection{Held-out Codeword Decodability (RQ2)}
\label{sec:heldout-dec}
Overall Recall and NDCG show that Grevo improves recommendation, but they do not reveal whether an evolved SID is itself easier for the recommender to generate. This distinction is important because Grevo selects edits using round-0 posterior diagnostics. So we perform a held-out codeword decodability test on Beauty: after retraining the recommender on the evolved index (round 1), we ask whether the new code is easier to predict in unseen user contexts.

For every edited item, we collect held-out validation/test histories—never used during candidate scoring—for which it is the next target. With the round-1 model and user history fixed, we score the original TIGER SID and the evolved SID as two alternative labels for that item via teacher forcing, reporting the negative log-likelihood (NLL) and the rank of each correct token. Because the model, user history, and target item are all held fixed, the SID is the only variable in each paired comparison; any improvement is attributable solely to the index edit and cannot reflect overfitting the round-0 posterior signal.

Figure~\ref{fig:heldout-dec} reports three complementary views of the result. Panel~(a) shows that all per-edit NLL points fall below the $y{=}x$ line. Panel~(b) confirms that the resulting NLL gains are uniformly positive with no regressions. Panel~(c) shows that evolved ranks concentrate near the top of the codebook while original ranks spread across a much wider range, motivating the logarithmic axis. Lower NLL means the model assigns higher probability to the full identifier; lower rank means fewer competing codewords during autoregressive decoding, reducing the risk of an early error that derails the remainder. Together, the three panels show that Grevo produces identifiers that remain genuinely more decodable after retraining.

\subsection{Hyper-Parameter Analysis (RQ3)}
\label{sec:settings}

\begin{figure*}[t]
    \centering
    \includegraphics[width=\textwidth]{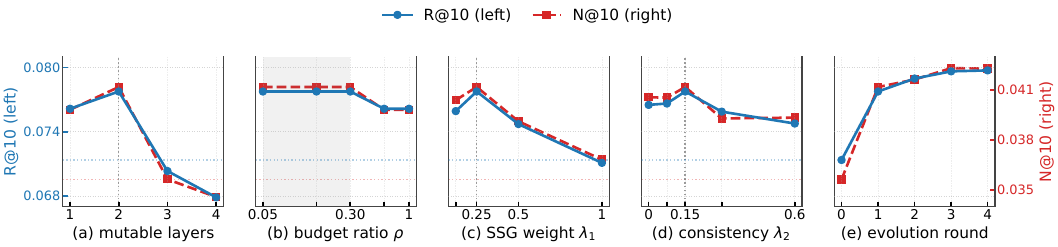}
    \caption{Hyper-parameter sensitivity and evolution-round analysis on Beauty. All five panels report absolute R@10 on the left axis and absolute N@10 on the right axis. Panels (a)--(d) vary one setting at a time, while panel (e) tracks cumulative evolution rounds.}
    \Description{Five panels show absolute R@10 and N@10 for mutable layers, budget ratio, SSG weight, consistency weight, and cumulative Grevo evolution rounds on Beauty.}
    \label{fig:hyperparam}
\end{figure*}

We further investigate the important hyper-parameters introduced by Grevo to facilitate future applications. All studies in this part probe Grevo's own internal knobs rather than compare against baselines, so we conduct them on the Beauty dataset with TIGER SID and vary one hyper-parameter at a time while holding the others at their default. Figure~\ref{fig:hyperparam} reports the absolute R@10 (left axis) and N@10 (right axis) for four hyper-parameters and then tracks both metrics over successive evolution rounds. We discuss each setting in turn.

\noindent\textbf{Number of mutable layers.}
An identifier has $K$ codebook layers, and Grevo can be restricted to reassign only a subset of them per round. Figure~\ref{fig:hyperparam}(a) shows that making two layers mutable gives the strongest result. Editing only the final layer leaves some local behavioral confusions unresolved, whereas opening more layers disrupts the stable higher-level structure of the identifier. So we use two mutable layers which preserves the semantic prefix while providing enough flexibility to repair the lower-level codewords that distinguish behaviorally similar items.

\noindent\textbf{Budget ratio $\rho$.}
The budget ratio $\rho$ caps the fraction of identifiers evolved per round through $\mathrm{Budget}=\lfloor\rho|\mathcal{I}|\rfloor$. Figure~\ref{fig:hyperparam}(b) shows a broad plateau at small-to-moderate budgets, while larger budgets do not improve downstream quality and eventually degrade it. We set $\rho=0.05$: it lies on this plateau but changes the fewest identifiers, reducing the cost of candidate evaluation and retraining while avoiding unnecessary edits. This pattern suggests that Grevo benefits from conservative, targeted updates rather than aggressive one-shot re-indexing.

\noindent\textbf{SSG weight $\lambda_1$.}
The weight $\lambda_1$ controls how strongly the semantic-grounding gain $\Delta r_{\mathrm{SSG}}$ contributes to the candidate score relative to the primary behavioral gain $\Delta r_{\mathrm{BSG}}$. Figure~\ref{fig:hyperparam}(c) peaks at $\lambda_1=0.25$. A smaller weight underuses semantic grounding, allowing behavior-driven edits that are less well supported by item semantics; larger weights increasingly suppress useful behavior-driven moves. We use $\lambda_1=0.25$ as the default, which best balances behavioral utility with semantic validity.

\noindent\textbf{Consistency weight $\lambda_2$.}
The weight $\lambda_2$ controls the cross-task consistency term $\Delta r_{\mathrm{cons}}$, which favors candidates on which BSG and SSG agree at every layer. Figure~\ref{fig:hyperparam}(d) shows a clear intermediate optimum at $\lambda_2=0.15$: without consistency, the two tasks can favor incompatible edits, while too much consistency over-penalizes behaviorally useful reassignments. We use $\lambda_2=0.15$, treating consistency as a lightweight regularizer that filters unreliable edits without overriding the main recommendation objective.

\noindent\textbf{Effect of evolution rounds.}
Beyond these per-round knobs, Grevo repeats \texttt{pretrain$\rightarrow$collect$\rightarrow$enumerate$\rightarrow$evolve$\rightarrow$retrain} for several rounds, and we study how downstream quality changes as rounds accumulate. Figure~\ref{fig:hyperparam}(e) tracks test R@10 and N@10 across rounds on Beauty with TIGER SID. Both metrics improve steadily during the early rounds, showing that retraining can turn successive index refinements into better recommendation quality. The gains then become progressively smaller, indicating that the remaining identifiers offer fewer high-value corrections after the most problematic assignments have been repaired. Performance has effectively converged after round~3, with later evolution producing little additional change.

\section{Conclusion}
\label{sec:conclusion}

We presented Grevo, a unified generative recommendation framework built from two components: a unified multitask model whose BSG and SSG tasks let a single recommender both recommend and absorb the tokenizer's role, and the evolutionary item indexing that improves SIDs by evolving the SID assignment rather than retraining a tokenizer. Instead of keeping a standing tokenizer coupled to the recommender in an unstable alternating loop, Grevo treats the tokenizer as a disposable initializer that bootstraps the SID assignment once with semantic priors and is then discarded, treating the SID assignment as an evolvable discrete variable and evolving a small, budgeted fraction of item identifiers after each training round. It reuses a unified generative recommendation model already trained on the current index as a posterior evaluator to score constrained candidate identifiers, then takes a coordinate-descent-like step along the direction of greatest alignment gain. The results highlight index quality as a critical factor in generative recommendation: directly optimizing the item index without a trainable tokenizer yields consistent gains across backbones and datasets, demonstrating strong potential as a practical and generalizable design principle. Looking forward, natural extensions include applying evolutionary indexing to streaming settings where item catalogs shift continuously, scaling to larger codebooks where more efficient candidate strategies are needed, and incorporating multi-modal embeddings as richer grounding signals for the SSG objective.

\bibliographystyle{ACM-Reference-Format}
\bibliography{evosid}

\end{document}